\documentclass[twocolumn,showpacs,prl,superscriptaddress,floatfix]{revtex4}

\usepackage{color} 
\usepackage{tabularx} 
\usepackage{epsfig}
\usepackage{amsmath} 
\usepackage{amssymb} 
\usepackage{graphicx}

\begin{document}
\title{Zero Droplet Stiffness Exponent $\theta$ is Revealed in Short Range Spin Glasses when Probed with Large Avalanches Induced by Long Range Interactions}

\author{Ferenc Pazmandi} 
\affiliation {San Carlos, CA 94070}

\author{Gergely T.~Zimanyi}
\affiliation {Department of Physics, University of California, Davis,
California 95616, USA}

\begin{abstract}
We probe the droplet excitations in short range spin glasses by adding a perturbative long range interaction that decays with distance as a power law: $J/r^{\sigma}$.  It is shown that if the power law exponent $\sigma$ is smaller than the spatial dimension $d$, the perturbation induces large scale avalanches which roll until they force the system to develop a pseudo gap in the excitation spectrum of the stabilities. This makes the perturbative long range interactions relevant for $\sigma < \sigma_c = d$.  The droplet theory predicts that the critical exponent $\sigma_c$ depends on the droplet stiffness exponent as $\sigma_c=d-\theta$.  Combining these two results leads to a zero stiffness exponent $\theta = 0$ in the droplet theory of short range spin glasses.

\end{abstract}


\maketitle

Spin glasses, and frustrated models in general, are relevant to a mesmerizingly wide variety of problems, such as random magnets, disordered electronic systems, neural networks, and even biological pathways.  In spite of the pressing need for an understanding of their physics, even after more than thirty years of study, the community remains profoundly split regarding the correct theoretical framework for short range spin glasses.  The two main approaches are the replica symmetry breaking (RSB) \cite{parisi:79,parisi:80,parisi:83,mezard:87}
and the droplet pictures.\cite{fisher:86,fisher:87,fisher:88,bray:86}  These theories can be distinguished, for example, by their predictions for the energy $E(L)$ required to reverse a cluster or droplet of spins characterized by a linear size $L$.  In the spin glass phase, the scaling form $E(L) \sim L^{\theta}$  defines the droplets' "stiffness exponent" $\theta$, which is positive in the droplet theory: $\theta>0$, but zero in the RSB theory: $\theta=0$.  

Whether the "stiffness" exponent is finite or not, has profound consequences for the essential physical mechanism of the spin glass transition. In the droplet picture the driving force of the glass transition is the emergence of the finite stiffness ($\theta>0$) of the droplets.  In contrast, in the RSB picture the stiffness remains weak ($\theta=0$) across the glass transition, which is driven instead by another mechanism: the divergence of barriers which fragment the free energy landscape. 

There are many other differences between the two theories' predictions, such as whether the droplets are fractal or not.  A recent attempt to bridge those differences introduced a hybrid approach, called the trivial-non-trivial (TNT) theory \cite{krzakala:00,palassini:00,katzgraber:01,marinari:00,palassini:00a,leuzzi:08}, which is built on the notion that different types of excitations of the same size $L$ may have different associated energy costs.  Droplet excitations with size $L$ cost an energy $E_{droplet}( L) \sim L^{\theta}$ whereas domain wall excitations of similar size cost another energy $E_{domain-wall}( L) \sim L^{\theta_{dw}}$.  Direct numerical measurements by the creators of the TNT approach demonstrated that the two stiffness exponents are different for the short-ranged Edwards-Anderson spin glass in $3d$: the droplet stiffness exponent $\theta$ is indistinguishable from zero, whereas the domain wall exponent is clearly positive $\theta_{dw}>0$.\cite{krzakala:00,palassini:00,katzgraber:01,marinari:00,palassini:00a,leuzzi:08}
 
However, all numerical works could be criticized for having been performed on systems with unconvincingly small sizes.  Thus, numerics is sometimes assigned secondary importance in the quest for the ultimate spin glass theory.\cite{moore:02}

The present Letter aims to contribute to this discussion by adopting the theoretical framework of the droplet theory and deriving a theoretical bound for the droplet stiffness exponent $\theta$ purely on theoretical grounds without resorting to numerics.  We probe the ground state of a short range spin glass by adding a long range interaction with a small amplitude, investigate the relevance of this perturbation, and compare our result with the considerations of the original droplet paper.\cite{fisher:86}  In particular, we determine how the distribution of local stabilities $P(\lambda)$ changes when the perturbative long range interaction is introduced.  We find that the perturbation induces avalanches which open a pseudo gap in the stability distribution $P(\lambda)$ and therefore become relevant if the exponent of the power law interaction $\sigma$ is below its critical value $\sigma_c$, equal to the spatial dimension $d$. 
 
The original droplet paper also determined the critical value of the $\sigma$ exponent and found that $\sigma_c$ depends not only on the spatial dimension $d$, but also on the stiffness exponent of the droplets: $\sigma_c=d-\theta$. \cite{fisher:86} The combination of these two exponent relations yield a zero droplet stiffness exponent, $\theta = 0$, purely on analytical grounds, which is our central result.  
      
In spin glasses, the local dynamics and energetics can be conveniently characterized in terms of stabilities which are the products of the local spins $S_i$ and their respective local fields $H_i: \lambda_i = S_i H_i = S_i \sum_j J_{ij} S_j$.  At zero temperature, all spins have positive stabilities in any locally stable state, including the ground state, and the values of these local stabilities describe the energy of one-spin excitations from that stable state.  For short range interactions,  the density $P(\lambda)$ of these stabilities assumes a finite value at zero stability: $P(\lambda=0) > 0$.\cite{binder:86} Thus, a fraction of the spins experience vanishing effective fields even in the ground state.  As such, they may be expected to be particularly sensitive to perturbations.

The energetics of the short range system is tested by adding perturbations characterized by weak random long range interactions whose amplitude decays with distance as a power law with an exponent $\sigma$: $J/r_{ij}^{\sigma}$, $J$ being a random number with a Gaussian distribution of mean $0$ and width $J$.\cite{kotliar:83}  We evaluate the relevance of this perturbation at zero temperature by analyzing whether the spin glass ground state is stable against avalanches induced by this perturbation.
  
The stability analysis starts by flipping a single spin after having added the long range perturbation and determining the average number of spins $N_{unstable}$ which get destabilized by this single flip.  Under a sequential spin-updating dynamics, one of these new unstable spins will flip in the next time step.  Thus, if a single spin flip destabilizes more than one additional spin on average, i.e., $N_{unstable} > 1$, then the spin flip starts an avalanche, or exploding chain reaction, causing a large scale destabilization of the ground state of the unperturbed short range spin glass.  If less than one spin is destabilized, i.e., $N_{unstable} < 1$, the avalanche dies out quickly, indicating that the ground state was stable against the perturbation.  Therefore, the avalanche stability criterion lies at exactly one newly destabilized spin per spin flip, i.e., $N_{unstable} = 1$. 
    
Fig. 1 illustrates the calculation of $N_{unstable}$.  Flipping a spin at site $i$ can destabilize spins at a distance $r$ from the site $i$ if the stabilities of the latter spins $\lambda(r)$ are smaller than their interaction with the flipped spin at site $i$: $\lambda(r) < J(r)=J/r^{\sigma}$.  For a short range spin glass, the stability distribution $P(\lambda)$ can be approximated at small stabilities by its value at zero stability $P(0) > 0$, yielding the following estimate for the average number of newly destabilized spins: 
\begin{equation}
N_{unstable}(r)=\int^{J(r)}_{0} d\lambda P(\lambda) \approx J(r) P(0)=P(0)J/r^{\sigma}
\end{equation}
\begin{equation}
N_{unstable}(R)=\int^{R}_{0} dr r^{d-1} N_{unstable}(r) \sim R^{d-\sigma}
\end{equation}
Here, Eq. (2) shows the summation of the unstable spins for distances up to the system size $R$. 
\begin{figure}
\includegraphics[width=8.0cm]{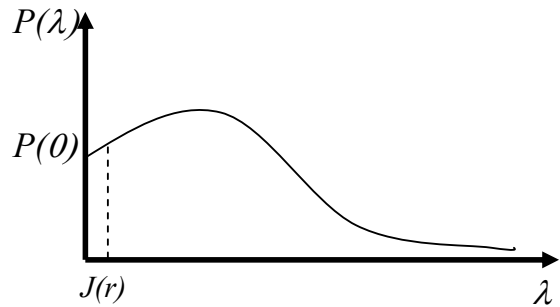}
\caption
{The calculation of $N_{unstable}(r)$. Upon the flipping of a single spin, the spins at distance $r$ with stabilities $\lambda<J(r)$ become unstable. Here P($\lambda$) is the distribution of the stabilities $\lambda$ for the short range spin glass}
\label{fig:Fig1}
\end{figure}
This estimate displays our central result clearly: for $\sigma < d$, a flip of a single spin destabilizes a large number of spins via the long range perturbative interaction: $N_{unstable}(R,\sigma < d) \sim R^{d-\sigma} >> 1$, irrespective of how small the amplitude of the perturbation.  This is so, because even though the strength of the long range interaction $J(r)$ diminishes with increasing distance $r$, but this effect is more than compensated by the number of spins impacted by $J(r)$ increasing with $r$, if $\sigma < d$. The subsequent flipping of these destabilized spins destabilizes an even more populous generation of spins, setting off a large avalanche.  

At the start of the avalanche, is it appropriate to analyze the number of spins, destabilized in subsequent steps, independently since the spatial structure of long range perturbation is not correlated in any way with the ground state configuration, created by the short range interaction.  While the rolling of the avalanches may start to develop such correlations, these should be of secondary importance as each spin flip destabilizes a large number of spins, thus sustaining the avalanche. 
 
Similar to the Efros-Shklovskii scheme for Coulomb glasses \cite{efros:75}, one expects that the avalanches will keep rolling until the system qualitatively transforms itself and eliminates the fuse of the avalanche, which is the large number of spins with low stabilities. Formally this is represented by $P(0)$ being greater than $0$ for the short range spin glass. The avalanches will roll until the average number of the destabilized spins $N_{unstable}$ is suppressed to or below $1$.  The minimal transformation of the system by which this stabilization can be achieved is that $P(\lambda)$ develops a power law form at small $\lambda$ stabilities: $P(\lambda) \sim \lambda^{\alpha}$. In the Coulomb Glass literature this process is called the opening a pseudo gap or soft gap. Indeed, a positive exponent $\alpha$ of the stability distribution reduces the number of destabilized spins:
\begin{equation}
N_{unstable}(r)=\int^{J(r)}_{0} d\lambda P(\lambda) \approx J(r)^{\alpha +1} 
\end{equation}
\begin{equation}
N_{unstable}(R)=\int^{R}_{0} dr r^{d-1} N_{unstable}(r) \sim R^{d-\sigma(\alpha +1)}
\end{equation}
From Eq.(4), this power law Ansatz for $P(\lambda)$ is capable of arresting the avalanches and stabilizing the system for $\alpha \ge \alpha_c=(d/\sigma) - 1$ by making the number of destabilized spins $N_{unstable}(R)$ non-divergent with the system size $R$. The corresponding opening of the pseudo gap in $P(\lambda)$ is illustrated in Fig. 2.
\begin{figure}
\includegraphics[width=8.0cm]{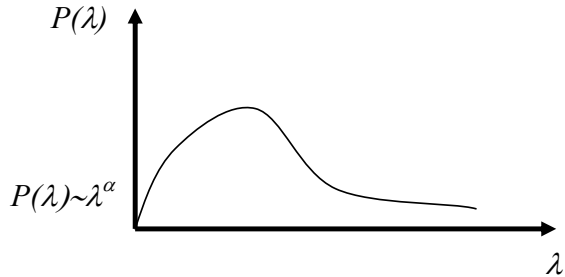}
\caption
{The $P(\lambda)$ distribution of stabilities develops a $P(\lambda) \sim \lambda^{\alpha}$ soft gap driven by the avalanches induced by the perturbative long range interaction.}
\label{fig:Fig2}
\end{figure}
Following the insights from the Coulomb gap literature, one expects that the system develops these large scale rearrangements only to the minimal degree necessary for stopping the avalanches.  In other words, the system drives itself only to the state of marginal, or critical stability.  This is achieved by the exponent $\alpha$ assuming the critical value $\alpha_c$:
\begin{equation}
\alpha=\alpha_c=(d/\sigma) - 1
\end{equation}
which saturates the Efros-Shklovskii bound.  In this sense, systems with long range interactions are similar to those which exhibit Self Organized Criticality (SOC).  Note that Eq.(5) holds only in the regime $d/2\le \sigma \le d$.  For $\sigma \le d/2$ the interaction has to be rescaled to ensure that the energy remains an extensive quantity. With the proper rescaling of the interaction $\alpha$ remains pegged at $\alpha(d/2) = 1$ for $\sigma \le d/2$, including the Sherrington-Kirkpatrick case of $\sigma=0$. For $\sigma \ge d$ the system remains effectively short ranged. Accordingly, no large scale avalanches form and no power law soft gap is needed for stability.   

In our earlier work on the hysteresis in the Sherrington-Kirkpatrick model we observed and reported the formation of such a pseudo gap with an exponent of $\alpha_{hyst} = 1$ for the states reached by a hysteretic history.\cite{pazmandi:99} This exponent $\alpha_{hyst} = 1$ is entirely analogous to the above $\alpha(\sigma=0) = 1$ result for the ground state.  In the work on hysteresis we also interpreted the saturation of the exponent bound as a sign of the system driving itself essentially to a Self Organized Critical state.

Returning to the central inquiry of the present work, Eq.(5) expresses the same criterion as Eq.(2) in the following sense.  For $\sigma < \sigma_c = d$, $\alpha$ becomes greater than zero and correspondingly the system is forced to respond to the long range perturbation by large scale avalanches, qualitatively rearranging the ground state and developing a pseudogap in $P(\lambda)$. In other words, for $\sigma < \sigma_c = d$ the long range interaction is a relevant perturbation.  

This relevant perturbation should be detectable through, for example, the low temperature behavior of many measurable quantities, such as the specific heat, the magnetic susceptibility, or, in electronic equivalents, the tunneling conductance. 

To complete our argument, we presented two separate analysis of the impact of a long range perturbation on a short range spin glass, based on (i) computing the number of unstable spins in avalanches induced by the perturbation, and on (ii) describing the development of a pseudogap, induced by the perturbation. Both of these arguments established that the perturbation by a long range interaction is relevant for the short range spin glass if $\sigma < \sigma_c = d$.  It is then recalled that the droplet theory predicted that the critical value $\sigma_c$ depends on the stiffness exponent for the droplets as: $\sigma_c = d-\theta$.  The combination of these two exponent relations suggests that the droplet theory $itself$ predicts that the stiffness exponent for droplets is zero: $\theta = 0$.
        
While our analytical considerations are self contained and do not require input or justification from numerical simulations, it is nevertheless reassuring that numerical results are in agreement with our conclusions.  Katzgraber and Young \cite{katzgraber:03}, for example, numerically studied the spin glass problem in $d=1$ with a long range interaction with exponent $\sigma$.  Working at $\sigma$ = 0.75, they determined the droplet stiffness exponent $\theta$ with great accuracy.  The droplet theory of \cite{fisher:86} predicted a droplet stiffness exponent $\theta = d-\sigma$ = 0.25 for this case.  Katzgraber and Young, however, found $\theta = -0.005$, clearly inconsistent with the droplet prediction, but consistent with our analytical result of $\theta = 0$, as well as with the TNT picture.  It is further noted that Ref. \cite{katzgraber:03} also studied the domain wall exponent and found it consistent with the TNT picture.  The very recent work by Leuzzi et al is also indicative of $\theta = 0$.\cite{leuzzi:08}    

We also mention that $P(\lambda)$ was explicitly analyzed numerically in a one dimensional version of the above long range interacting model in a related recent paper.\cite{boettcher:08a}  While $P(\lambda)$ exhibited a strong suppression approaching $\lambda = 0$ for $\sigma < 1$, the data did not compel the conclusion of $P(\lambda=0) = 0$.  However, we believe that further work is in order here, as measuring a small exponent in the presence of large finite size effects requires a careful analysis.  Our ongoing work on this issue will be reported elsewhere.\cite{gonzalez:09}

The stiffness exponent being zero also has support from papers directly simulating short range spin glasses.  Kisker et al. studied the coarsening dynamics of the $3d$ Edwards-Anderson Spin Glass and concluded that the scaling analysis of the autocorrelation function and correlation length in particular were suggestive of the stiffness exponent $\theta = 0$.\cite{kisker:96}  Their data were best fit with a droplet energy cost which depended logarithmically on the linear size: $E_{droplet}(L) \sim ln(L)$. The vanishing of the droplet stiffness exponent $\theta$ was also concluded from equilibrium studies of system size excitations.\cite{krzakala:00,palassini:00,katzgraber:01,marinari:00,palassini:00a,leuzzi:08}
 
One could propose to resolve the above conundrum by arguing that the original exponent relation derived in \cite{fisher:86} was only a lowest order calculation and it needs to be reanalyzed in light of the importance of the avalanches.  However, this only proves our point once again.  The calculation in \cite{fisher:86} was carried out assuming that the system was stable against perturbations.  In stable phases, such lowest order perturbative calculations should be appropriate. Therefore, if it is argued that the avalanches rearrange the state of the short range spin glass to such a degree that a higher order, or even a re-summed infinite order perturbation calculation is necessitated because e.g. $P(\lambda)$ develops a soft gap, that is essentially a recognition that the long range perturbations were indeed relevant, proving our point once more.  

Before closing we mention that no explicit analysis was given above for the domain wall stiffness. Deriving the analogous exponent relation and the consequences of the pseudogap formation for domain walls is the target of ongoing work.   

In conclusion, we have analyzed the stiffness exponent $\theta$ of the short range Ising Spin Glass by applying the "probe" of an additive perturbative long range interaction. We found that the long range interactions induce large avalanches in the system because of the large number of low-stability spins in the short range Ising Spin Glass. These avalanches roll until the system develops a pseudo gap at low energies, characterized by a power law distribution of the stabilities $P(\lambda) \sim \lambda^{\alpha}$, where $\alpha = (d/\sigma) - 1$ for $d/2 \le \sigma \le d$ and $\alpha = 1$ for $\sigma \le d/2$.  This large scale reorganization makes the long range interaction a relevant perturbation for all $\sigma \le \sigma_c = d$ exponent values.  Finally, comparing this critical exponent with the prediction of the droplet theory of $\sigma_c = d- \theta$, one concludes that combining the logic of the droplet theory with the notions of avalanches and pseduogaps points to the stiffness exponent being $\theta = 0$ in the short range Ising Spin Glass.  While the vanishing of the droplet stiffness has consequences for the droplet theory, it is consistent with the RSB and TNT theories, both of which are consistent with such a weak stiffness.

\begin{acknowledgments} 

We acknowledge insightful discussions with H. Katzgraber, V. Dobrosaljevic, D. Huse, K. Pal, and A.P. Young.  
\end{acknowledgments}

\vspace{-0.4cm} 
\bibliography{refs}

\end{document}